# Utilizing Concept Drift for Predicting the Time Lag of Policy Interventions: The Case of the COVID-19 Pandemic

Lucas Baier[1], Niklas Kühl[1,2], Jakob Schöffer[1], Gerhard Satzger[1,2]
[1]Karlsruhe Institute of Technology (KIT)
[2]IBM

**Abstract:** As a reaction to the high infectiousness and lethality of the COVID-19 virus, countries around the world have adopted drastic policy measures to contain the pandemic. However, it remains unclear which effect these measures, so-called non-pharmaceutical interventions (NPIs), have on the spread of the virus. In this article, we use machine learning and apply drift detection methods in a novel way to predict the time lag of policy interventions with respect to the development of daily case numbers of COVID-19 across 9 European countries and 28 US states. Our analysis shows that there are, on average, more than two weeks between NPI enactment and a drift in the case numbers.

## 1   Introduction

Within just a few months in early 2020, the COVID-19 disease caused by a novel coronavirus has evolved into a global pandemic. In order to fight the spread of the pandemic, drastic policy measures with far-reaching implications for basic civil rights have been passed. Politicians have to carefully assess these so-called non-pharmaceutical interventions (NPIs) and gauge their necessity and precise timing. Typical NPIs observed across many nations include a lockdown of public life and the closure of schools. Despite consensus in the scientific community on the general effectiveness of a combination of NPIs to mitigate the progression of COVID-19, their concrete timing, duration, scope, and effect have controversially been discussed in public.

In this article, we introduce a strategy to measure and predict the time lag between the introduction of policy measures and a significant change in the subsequent data, and we illustrate it with NPI and case data from the COVID-19 pandemic. We propose to utilize the kernel theory of *concept drift detection* (Gama et al., 2014) to identify significant changes in the development of the number of infected persons. Concept drift detection is a machine learning technique usually applied to detect changes—so-called *drifts*—in the data-generating environment and trigger retraining of the corresponding model. In the field of Information Systems (IS), concept drift detection has been applied to, e.g., data stream analysis (Brzezinski & Stefanowski, 2017), business process mining (van Zelst et al., 2019), or innovation monitoring (Mirtalaie et al., 2017). We apply this technique in a novel way, as we do not use it to improve prediction performance on a data stream but to detect significant changes in variables of interest (e.g., COVID-19 case numbers). Specifically, these variables of interest are monitored for drifts that *might* be related to previous interventions, e.g., of political or medical nature. Following a Design Science Research (DSR) approach, we build an artifact applying concept drift detection for this purpose and illustrate and evaluate it for the pandemic use case.

We show that concept drift detection can be effectively applied to detect sudden changes in a target variable—in this particular case, changes of COVID-19 infection numbers. We base our analysis on data from 9 European countries and 28 US states obtained between 22-01-2020 and 12-05-2020, and we discover that there are on average more than two weeks between the enactment of an NPI and a drift in the data. We continue to analyze the predictive power of each NPI as well as a set of additional features with respect to forecasting the timing of drifts in case numbers.

The remainder of this article is structured as follows: Section 2 introduces the underlying research design, before Section 3 covers related work and introduces the kernel theory of *concept drift*. Section 4 explains the design of our artifact which is subsequently evaluated in Section 5. With the results at hand, we discuss our findings in Section 6. Section 7 concludes our work.

## 2 Research design

As an overall research design, we choose DSR, as it allows to consider the design-related tasks necessary when building IT artifacts (March & Smith, 1995). Moreover, it has proven to be an important and legitimate paradigm in IS research (Gregor & Hevner, 2013).

We design an artifact capable of measuring the time lag between interventions (e.g., policy measures) and a drift in temporal data. Our artifact is best described as a *method*, as it consists of "actionable instructions that are conceptual" (Peffers et al., 2012, p. 401). By designing and applying our method, we aim to show its *feasibility* (Pries-Heje et al., 2008) as our key evaluation criterion. As a result, our research contributes to the two dimensions of DSR projects (Gregor & Jones, 2007), namely generalizable design knowledge ("knowledge on how to *build* the artifact") as well as a significant impact within the field of application ("knowledge resulting from the *use* of the artifact"). Regarding the former, we design an artifact utilizing concept drift algorithms to detect significant changes in a target variable like the development of case numbers during a pandemic. Therefore, we inform the design of information systems by demonstrating how the kernel theory of *concept drift* can be applied in novel ways. In terms of knowledge contribution according to Gregor & Hevner (2013), the artifact is best described as an *improvement,* as it applies a known solution (concept drift detection) to a novel problem. The overview of our chosen research design as well as the integration into the existing DSR literature is depicted in Table 1.

*Table 1. Overview of DSR project characteristics.*

| **Real-world problem** | Measure time lag between policy interventions and a drift in COVID-19 case numbers |
|---|---|
| **Kernel theory** | Concept drift detection |
| **Artifact type** (**Peffers et al. (2012)**) | Method |
| **Evaluation objective** (**Pries-Heje et al. (2008)**) | Feasibility |
| **Evaluation type** (**Venable et al. (2016)**) | Technical risk & efficacy |
| **Contribution** (**Gregor & Hevner (2013)**) | Improvement; application of known solution (drift detection) to novel problem |

# 3 Related work

## 3.1 *Concept drift*

Supervised machine learning fits a mathematical function to map input features to a corresponding, to-be-predicted, target. This function is usually learned by considering historical data as training data. The resulting model can continuously create value when deployed in information systems and delivering ongoing predictions on continuous data streams of new incoming data. However, data streams usually change over time, which also leads to changes in the underlying probability distribution (Tsymbal, 2004). This challenge of changing data over time is usually described as *concept drift* in computer science (Widmer & Kubat, 1996). A concept *p(X,y)* is defined as the joint probability distribution of a set of features *X* and the corresponding label *y* (Gama et al., 2014). The change of a concept over time can be expressed in a mathematical definition as follows:

$$\exists X: p_{t_0}(X, y) \neq p_{t_1}(X, y).$$

Therefore, concept drift is defined as the change in the joint probability distribution between two time points $t_0$ and $t_1$. This change may entail that the machine learning model built on previous data in $t_0$ is no longer suitable for making predictions on new incoming data in $t_1$. Thus, the occurrence of concept drift requires the application of countermeasures, e.g., the frequent retraining of the underlying machine learning model.

Changes in the incoming data stream can be triggered by a multitude of different internal or external effects. In general, it is intractable to measure all possible confounding factors—which is why these factors cannot be integrated directly into the machine learning model. Those factors are often considered as hidden context of a predictive model (Widmer & Kubat, 1996). The phenomenon of concept drift is usually classified into the following categories (Žliobaitė, 2010): *Abrupt or sudden* concept drift where data structures change very quickly (e.g., a sudden drop in airline traffic during COVID-19), *gradual and incremental* concept drift (e.g., change in customers' buying preferences), or *seasonal and reoccurring* drifts (e.g., ice cream sales in summer). The more fine-grained taxonomy of Webb et al. (2016) also considers other factors, e.g., the magnitude of the drift.

Different approaches for the handling of concept drifts have been proposed: In general, the adaptation strategies for the machine learning model can be split into *blind* and *informed methods* (Gama et al., 2014). Blind adaptation strategies adapt or retrain the prediction model without any explicit drift detection strategy, usually in fixed time intervals (e.g., every month). In contrast, informed methods rely on explicit concept drift detection algorithms which are able to detect concept drifts and trigger a corresponding warning. These drift detection algorithms can further be classified into three categories (Lu et al., 2018): The first category, *error rate-based drift detection*, tracks changes by analyzing the error rate of the prediction model. If the error rate changes significantly over time, a drift alarm is triggered. The second category, *data distribution-based drift detection*, measures the dissimilarity between the distributions of historical data and more recent data. The third category, *multiple hypothesis test drift detection*, combines several techniques of the previous two categories. In general, most approaches belong to the first category (Lu et al., 2018), with Page-Hinkley Test (Page, 1954) and ADWIN (Bifet & Gavalda, 2007) being two of the most popular algorithms. The Page-Hinkley Test (PHT) works by continuously monitoring an input variable (e.g., the input data or the prediction accuracy). As soon as the variable differs significantly from its historical average, a change is flagged. ADWIN, in contrast, is a change detector which relies on two detection windows. As soon as the means of those two windows are distinct enough, a change alert is triggered, and the older window is dropped.

In light of the contribution of this article, it is worth noting that all aforementioned methods are usually applied to detect drift in the joint probability distribution of a set of features and a *target variable*. This information is then applied to adapt a supervised machine learning model. In this work, we apply drift detection to identify changes in a *target variable* only.

### 3.2 *Measuring the spread of pandemics*

Modeling and predicting the development of infectious diseases can be performed with different tools, such as compartmental models, agent-based models, or time series and machine learning models (Nsoesie et al., 2014). Compartmental models, such as SIR (Schoenbaum, 1924) or one of its variations, work by dividing the population in compartments based on the disease state (such as susceptible, infected, and immune/dead) and computing rates at which individuals switch between compartments (McCluskey, 2010), typically using Markov chains. Agent-based approaches model the behavior of individuals and their interactions and thereby allow for analyzing the overall transmissions. In contrast, time series and machine learning models rely on past case data and predict future values on that basis.

Regarding the spread of COVID-19, variations of the SIR model have been applied to model and predict the transmission in Hubei and other regions of China by integrating population migration data (Yang et al., 2020). Similar work has been done for India (Pandey et al., 2020). Other approaches use dynamic SIR models to account for changing reproduction numbers following NPIs for different countries (Fanelli & Piazza, 2020). Time series and machine learning approaches are also widely applied, e.g., for predicting the Italian case numbers with exponential curves (Remuzzi & Remuzzi, 2020), or by applying exponential smoothing models (Petropoulos & Makridakis, 2020). The impact of case importation from different areas on transmission rates can be investigated with generalized linear models (Kraemer et al., 2020). Furthermore, neural network-based methods such as LSTMs have been trained on earlier outbreaks of SARS and have been applied to predict the spread of COVID-19 (Yang et al., 2020). An agent-based model originally developed for flu prediction has been used for the analysis of the spread in Singapore (Koo et al., 2020).

## 4 Artifact design

The design of our artifact is informed by knowledge from the field of concept drift detection. As stated in Section 3.1, we use drift detection on a *target variable* to identify a substantial change, potentially caused by prior interventions, among others. We select the Page-Hinkley Test (PHT) as a representative for our kernel theory of concept drift detection, as it is one of the most widely used concept drift detectors (Mitrovic et al., 2018). Based on the PHT, we build an artifact which takes as input a time series data stream as well as corresponding predictions and calculates drifts in the data. The overview of our artifact's workings is depicted in Figure 1.

The artifact calculates the time lag between NPIs and the detected drift for one country/state (output). As input, our artifact requires time series data of infected cases from the respective region as well as a prediction model trained on previous data instances. We fit an exponential smoothing model (Holt, 2004) with a seasonal component and a multiplicative trend. This allows for better modeling the spread of the disease in many different countries/states since it enables us to incorporate *exponential* rather than solely linear trends. Optimal values for the smoothing parameters are determined through grid search.

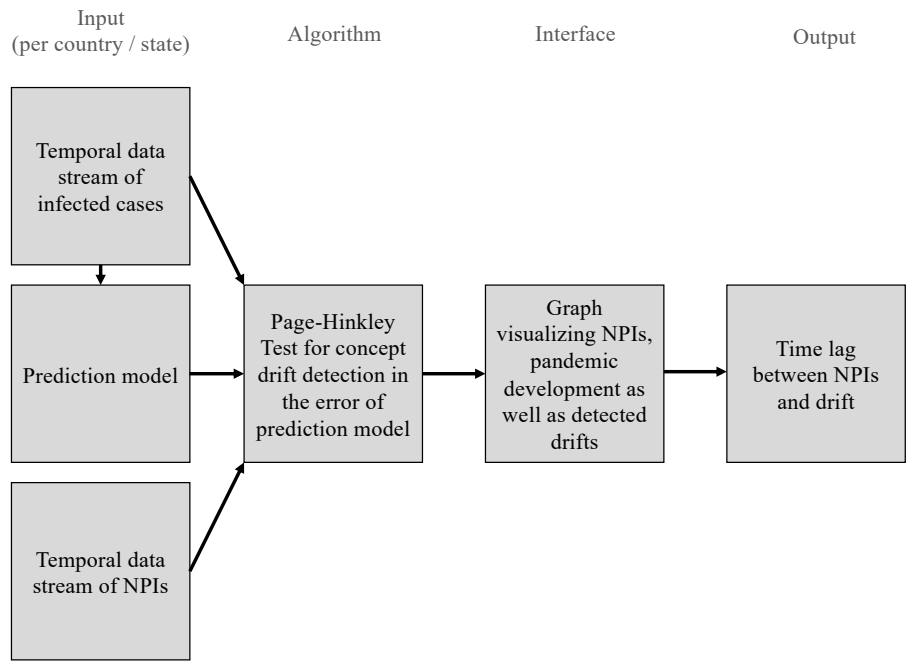

Figure 1. Overview of the presented artifact.

As an example, Figure 2 shows the daily case numbers (in grey) as well the predictions (in blue) of the exponential smoothing model for Spain. The figure confirms that the prediction model is well able to accurately represent the exponential development of the case numbers during the unrestricted spread of COVID-19.

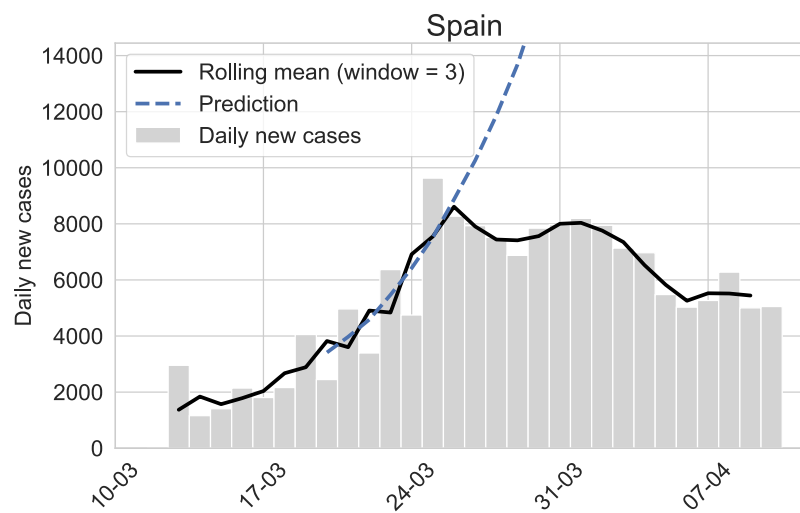

Figure 2. Daily new case numbers as well as corresponding predictions for Spain.

Subsequently, we compute the predictions for future case numbers and compare those predictions with the actual reported case numbers. To that end, we use the symmetric mean absolute percentage error (SMAPE) (Tofallis, 2015). If the error is small, the true and predicted case numbers are similar, which suggests that the pandemic evolves as expected based on the prediction model trained on historical case numbers. A large error, however, indicates that the pandemic is not evolving as predicted. We identify those dates with a significant discrepancy between true and predicted numbers—a concept drift—by applying the PHT on the SMAPE metric. We assume that in case we have identified a drift, it is associated with a change in the general trend of the case numbers. Finally, we measure the time lags between NPI enactment and the detected drift. The detected drift, the development of the case numbers as well as the enactment of NPIs are then visualized within a holistic graph (Figure 3). More details on the precise implementation can be found in the appendix in Table 4.

## 5 Artifact evaluation

We instantiate our artifact for evaluation purposes with data from the COVID-19 pandemic. We use the daily new infections with COVID-19 between 22-01-2020 and 12-05-2020 as data input. The data is provided by the Center for Systems Science and Engineering at Johns Hopkins University (JHU CSSE, 2020), which also forms the basis for the well-known COVID-19 dashboard (Dong et al., 2020). The data set does not only contain the worldwide case numbers at the country level but also provides information at the county and state level for certain nations such as the US.

We include 9 European countries in our analysis: Austria, Belgium, Germany, Italy, Norway, Spain, Sweden, Switzerland, and UK. Regarding the US, we refrain from a country-wide analysis as the occurrence of infections differs widely across the country. Instead, we perform a more detailed analysis at the state level. However, meaningful case predictions for states with few reported COVID-19 cases are difficult to compute. Therefore, we restrict our analysis to US states with more than 10,000 cumulated cases as of 13-05-2020. This leaves 28 states to be included, with New York, New Jersey, Illinois, Massachusetts, and California being the US states with most COVID-19 cases and Mississippi the least affected one. In total, we consider 37 countries or states in our analysis. Europe NPI data are sourced from Flaxman et al. (2020), US NPI data from Keystone Strategy (2020). We gathered data on mask wearing enactment dates for each country/state individually from national news outlets.

For this analysis, we consider five types of NPIs: *gathering restrictions, social distancing measures, closure of schools, lockdowns* (closure of non-essential services), and *mask wearing*. However, it should be noted that, even though named similarly, the execution of NPIs may differ significantly between countries/states. In "lockdowns", e.g., citizens in Italy were not allowed to leave their apartment for outdoor physical activities whereas in Germany they could still go for a walk with one person from a different household (Deutsche Welle, 2020). This type of lockdown is yet completely different from the lockdown in Wuhan, where public transport was shut down and inhabitants were only allowed outside for grocery shopping a few days a week (Graham-Harrison & Kuo, 2020). Additionally, the considered US states have not issued lockdown orders to date, which is why we use the closure date of non-essential services instead.

In the following, we instantiate of our artifact with an evaluation of the time lag between NPIs and detected drift. Furthermore, we investigate the possibility to predict the timing of the drift based on the enactment date of the NPIs and other relevant features.

### 5.1 Analysis of time lag between NPIs and drift

Figure 3 illustrates the development of daily cases in both Italy and New York since the end of February 2020. Furthermore, we show the relative spread of the pandemic per country/state by computing the number of deaths in relation to its population size. We consider the number of deaths as a reference point since this number is assumed to be a more reliable indicator for the state of the pandemic than the number of infected persons (Flaxman et al., 2020). Therefore, we identify the date with one COVID-19 death per one million inhabitants (red vertical line). For instance, this date is 03-03-2020 for Italy (60 million inhabitants, 60 cumulated deaths are counted on 03-03-2020). The different NPIs are indicated by green vertical lines, e.g., in Italy school closures on 05-03-2020, gathering restrictions and social distancing measures on 09-03-2020, and the following lockdown on 11-03-2020. In blue, the prediction model is shown, which is fitted with the data before 12-03-2020. The drift as indicated by the PHT takes place on 22-03-2020. This allows to compute the difference between the drift date and respective NPIs, e.g., the difference between school closure and drift is 17 days.

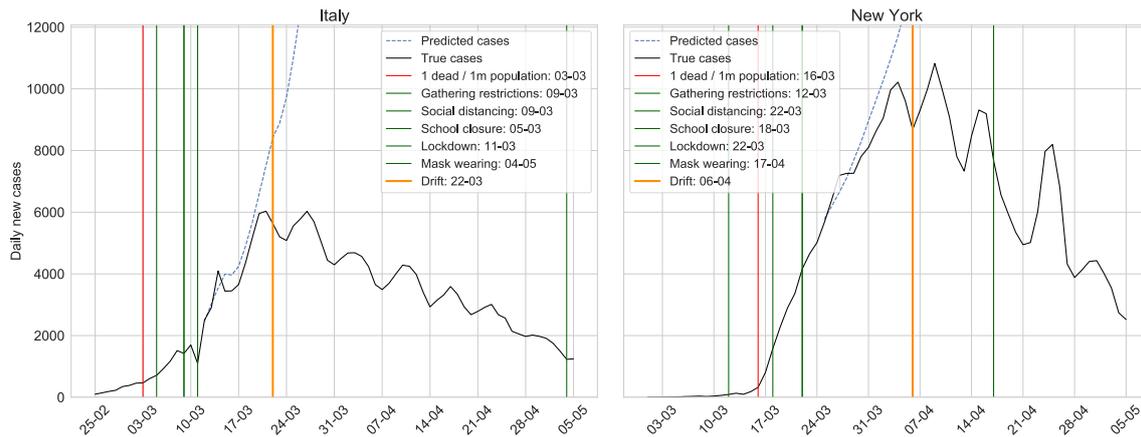

*Figure 3. Development of case numbers, prediction models, NPIs, and different drifts in both Italy (left) and New York (right).*

For every country/state included in our data set, we perform the same computation as depicted in Figure 3. As a result, we obtain the time lags between the NPI implementations and the drift. The mean and standard deviation for these time lags (in days) across all countries/states are depicted in Table 2.

*Table 2. Overview of time lags between NPIs and drift.*

| NPI | Time lag between NPI and detected drift [days] | Standard deviation [days] |
|---|---|---|
| Gathering restriction | 16.47 | 5.57 |
| School closures | 16.08 | 3.05 |
| Social distancing | 13.42 | 6.62 |
| Lockdown | 8.94 | 6.05 |
| *(Mask wearing* | *-44.48* | *29.61)* |

On average, the NPI *gathering restriction* is the first NPI taken during the course of this pandemic. The mean time between this NPI and the detected drift in the data is 16.47 days. While the first four NPIs were all established early in the pandemic and actually precede the detected drift (resulting in positive mean times in Table 2), the NPI of mask wearing was on average introduced 44 days *after* the detected drift. This is indicated by a negative time difference between NPI and drift date in Table 2. Specifically, even for *each* individual country/state, the mask wearing NPI was introduced after the detected drift—with a time lag ranging from 9 (New Jersey) to 101 days (Switzerland). For a detailed overview, see Table 5 in the appendix.

## *5.2 Predicting the NPI time lag*

As indicated by the standard deviations in Table 2, there are differences in the time lag of NPIs across various countries and states. To see whether the drift point might be predictable, we build a predictive model across all countries/states. As dependent variable, we choose the time lag between the date of one death per one million inhabitants and the drift in case numbers. We need this relative measure because the timing of the enacted NPIs as well as the status of the pandemic varies per country/state, and a purely date-based time series analysis would not account for this. A short time lag indicates that a country/state has reached a drift point in COVID-19 cases early. Note that this time lag is different compared to the values introduced in Table 2 due to the different reference points.

We collect a set of features from different categories which we (a) hypothesize to be predictive of the spread of the pandemic and (b) are publicly available. The decisiveness of a country/state is represented by the *reaction time*, which we define as a feature which measures how early a country/state reacted with their NPIs relative to the specific development of the pandemic. Note that we do not include mask wearing as NPI in this regression analysis: In every country/state, the mask wearing NPI was introduced *after* the detected drift (negative time difference between NPI and drift in Table 2). Therefore, this NPI cannot possibly explain the different time lags across countries/states.

We explain the computation of the reaction time in the following: For instance, Italy has introduced school closures (05-03-2020) two days *after* the relative death threshold of one death per one million inhabitants (03-03-2020), resulting in a reaction time of 2 days. Another example is Austria, which has reached one death per one million inhabitants on 21-03-2020. Gathering restrictions in Austria were already introduced on 10-03-2020, which means that this action was taken 11 days *before* the relative death threshold, resulting in a reaction time of -11 days. We compute the reaction time for all countries/states as well as for the NPIs *gathering restrictions*, *social distancing*, *closure of schools*, and *lockdowns*. Since not all countries/states have imposed all four NPIs (e.g., Sweden did not introduce a lockdown), we remove those instances for the following analysis, leaving us with 28 countries or states[*]. Since we assume a relationship between infections and population density, we collect the *population density per km²* as well as the *share of urban population*. General economic metrics of interest are represented by *GDP per capita in $* and the *Gini coefficient of income distribution*. Furthermore, we gather *healthcare expenditure per capita in $* and the *number of hospital beds per 100,000 inhabitants* to approximate the quality of the health care system. Climate is considered by including the *average temperature in March 2020*.

This data gathering process leads to a data set with 28 observations (countries or states) and 13 features each (4 reaction times plus 9 features related to metadata). This is a data set which is difficult to process for most learning algorithms, since the number of features is rather large compared to the number of data observations (Bellman, 1962).

---

[*]Austria, Belgium, Germany, Italy, Norway, Spain, Switzerland, United Kingdom, New York, New Jersey, Illinois, Massachusetts, California, Michigan, Texas, Florida, Georgia, Connecticut, Louisiana, Virginia, Ohio, Indiana, Colorado, North Carolina, Wisconsin, Alabama, Missouri, Mississippi

Therefore, we decide to apply a feature reduction method to improve the predictive performance. Instead of applying a feature selection method before model fitting, we rely on Lasso regression (Tibshirani, 1996), which performs feature reduction as well as model fitting simultaneously. In contrast to other shrinkage methods such as Ridge regression, Lasso not only shrinks the estimated coefficients of the linear regression but also forces some coefficients to be exactly equal to zero (James et al., 2013). This effect is achieved by applying the $L_1$-norm for penalizing the coefficients. Therefore, Lasso effectively performs parameter selection. The amount of shrinkage or regularization is controlled by the shrinkage parameter $\lambda$. Larger values of $\lambda$ lead to stronger regularization, thereby leading to smaller coefficients and more coefficients which are effectively equal to zero. Selecting a good value for $\lambda$ is critical and largely influences the quality of the predictive model. An appropriate value is often determined by performing cross-validation (Tibshirani, 1996).

For determination of the shrinkage parameter, we perform a random search (Bergstra & Yoshua, 2012) based on a set of 500 random values taken from a uniform distribution in the range of [0,5]. Since we are confronted with a data set limited in size, we decide to apply a nested cross-validation with five outer folds and three inner folds for evaluation (Cawley & Talbot, 2010). This method allows us to optimize the shrinkage parameter, while at the same time preventing overfitting on the test data. We use the three inner folds for the optimization of $\lambda$, which effectively splits those folds into training and validation data. The five outer folds serve as unseen test data and are used for estimating the prediction performance of the model. Before fitting the Lasso model, we standardize the input data to have zero mean and unit variance. Based on this setup, $\lambda$ takes the following five values after parameter optimization: 1.17, 1.59, 1.29, 0.79, 1.05.

Based on the nested cross-validation setup, we can compute a prediction for every data instance included in the data set. Based on this setup, we achieve a predictive performance with a mean absolute error of 3.02 days on the respective test set (RMSE = 3.72, $R^2$ = 0.53). The prediction results are plotted in Figure 4. Those results clearly illustrate that the model is able to predict the timing of the drift in case numbers.

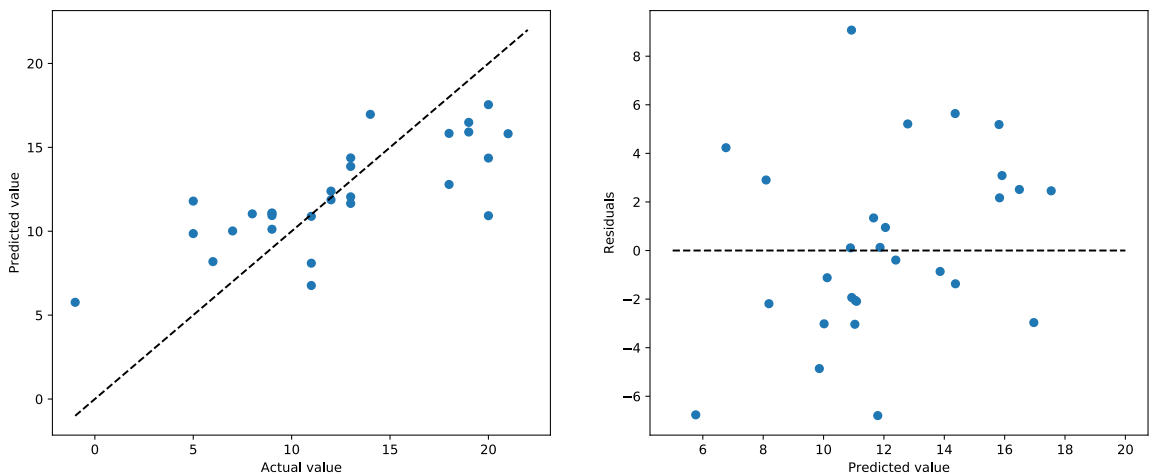

*Figure 4. Prediction results of Lasso model.*

Besides measuring the predictive power of the model, we can also analyze the importance of different input features by averaging the weights of the Lasso coefficients on the five outer folds. As expected, the model efficiently performs a feature reduction by determining only six non-zero weights (five input features plus intercept). Therefore, model size is reduced from 14 to six features only. All non-zero coefficients are depicted in Table 3.

*Table 3. Average Lasso regression coefficients across different folds.*

| Feature | Coefficient |
|---|---|
| Gini coefficient in 2018 | 0.022 |
| Household size in 2018 | 0.011 |
| GDP per capita in 2018 [US $] | -0.018 |
| Reaction time *gathering restrictions* [days] | 0.322 |
| Reaction time *school closure* [days] | 3.001 |
| Intercept | 12.284 |

Furthermore, we plot the average of all features (depicted by the height of the gray bars) with non-zero coefficients as well as the estimated coefficient values within each of the five different models fitted (dots in different shades of blue). This is depicted in Figure 5. An analysis reveals that the estimated coefficients are stable across the different folds (dots in the figure).

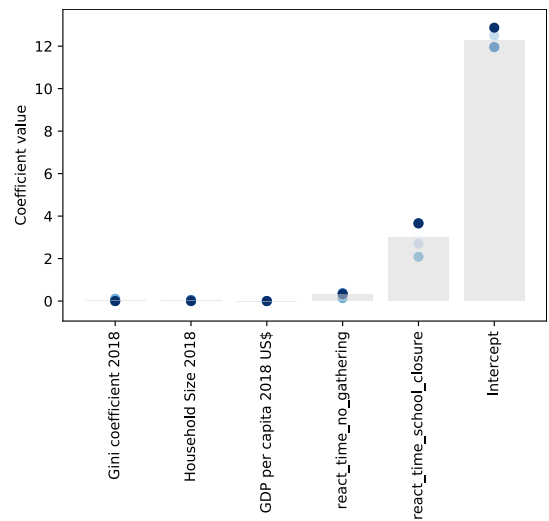

*Figure 5. Average coefficient values and value per fold.*

Due to the standardization of the input features, the size of the coefficients reveals their importance for predicting the target (time lag between the date of one death per one million inhabitants and the drift in case numbers). Based on Figure 5, we may conclude that the reaction time for gathering restrictions as well as the reaction time for school closures are the most important variables for computing a prediction. This behavior is also consistent across all five folds. However, note that this analysis does not imply any causal relationships (Hernán et al., 2019) between our predictors and the drift in case numbers—the objective of this work is solely to investigate whether a prediction of this time point (i.e., the drift) is possible. In contrast, a causal analysis for the effect of different NPIs on the drift date is very difficult in this setup, as different NPIs were often introduced together. Furthermore, by applying a Lasso model, we may remove variables from the modeling process which might have a causal relationship with the drift date.

## 6   Conclusion

The COVID-19 pandemic poses many challenges to politics and society. Especially the adopted policy measures to control the spread have been subject to heated debates. While we consciously refrain from contributing to the debate around the effectiveness of specific non-pharmaceutical interventions (NPIs), we have motivated and introduced an approach towards predicting the time lag between interventions and a significant change in a variable of interest. Specifically, we propose to utilize the kernel theory of *concept drift detection* (Gama et al., 2014) to measure the time difference between the introduction of certain NPIs and a significant change (so-called *drift*) in the number of infected cases. To evaluate our approach, we instantiate the proposed artifact based on actual data from the COVID-19 pandemic from spring 2020 with the goal to evaluate the general feasibility (Pries-Heje et al., 2008). Our analysis shows that the detected amount of time between the first adopted NPI and drift amounts to 16 days on average across 37 countries/states. Additionally, we show that we can predict the timing of drifts in case numbers based on a set of reaction times regarding different NPIs as well as other features. By applying a Lasso Regression model, we can achieve a mean absolute error of 3.02 days.

The work at hand contributes to the body of knowledge in two meaningful ways: First, we generate generalizable design knowledge in a DSR project. Our research demonstrates the successful application of concept drift detection as a design principle and illustrates how it can be applied in novel ways, thus informing the design of innovative information systems. Specifically, we apply drift detection to identify significant changes in a *target variable* and subsequently measure the time lag between this drift and the prior enactment of interventions—and we apply the artifact to the case of NPI and infection data from the COVID-19 pandemic, our area of application.

# Appendix

*Table 4. Implementation details of artifact.*

| Component | Details |
|---|---|
| Prediction model | Exponential smoothing model with additive seasonal component and multiplicative trend. This allows to model exponential trends. |
| Seasonality | Seven-day seasonality, as many countries/states exhibit a weekly pattern in their reporting of case numbers. |
| Grid search | We determine the optimal values for the model smoothing parameters for level, slope, and season by performing a grid search in the range (0.1, 0.2, … 0.9) and testing those values on the three days following the training set. |
| Training data | We use all case numbers up to seven days after the first NPI as training data. |
| PHT parameters | Threshold = 0.3<br>Minimum number of instances = 3 |

*Table 5. Overview of the mask wearing NPI analysis.*

| Country/State | Detected drift | Introduction of mask NPI | Difference in days between drift and introduction of mask NPI |
|---|---|---|---|
| Austria | 2020-03-28 | 2020-04-14 | 17 days *after* detected drift |
| Belgium | 2020-03-31 | 2020-05-04 | 34 days *after* detected drift |
| Switzerland | 2020-03-27 | 2020-07-06 | 101 days *after* detected drift |
| Germany | 2020-03-30 | 2020-04-27 | 28 days *after* detected drift |
| Spain | 2020-03-29 | 2020-05-21 | 53 days *after* detected drift |
| United Kingdom | 2020-04-06 | 2020-06-15 | 70 days *after* detected drift |
| Italy | 2020-03-22 | 2020-05-04 | 43 days *after* detected drift |
| California | 2020-04-05 | 2020-06-18 | 74 days *after* detected drift |
| Connecticut | 2020-03-31 | 2020-04-20 | 20 days *after* detected drift |
| Illinois | 2020-04-01 | 2020-05-01 | 30 days *after* detected drift |
| Maryland | 2020-03-30 | 2020-04-18 | 19 days *after* detected drift |
| Massachusetts | 2020-03-31 | 2020-05-06 | 36 days *after* detected drift |
| Michigan | 2020-03-31 | 2020-04-26 | 26 days *after* detected drift |
| New Jersey | 2020-04-01 | 2020-04-10 | 9 days *after* detected drift |
| New York | 2020-04-06 | 2020-04-17 | 11 days *after* detected drift |
| North Carolina | 2020-03-30 | 2020-06-26 | 88 days *after* detected drift |
| Pennsylvania | 2020-03-29 | 2020-04-19 | 21 days *after* detected drift |
| Rhode Island | 2020-03-30 | 2020-04-18 | 19 days *after* detected drift |
| Texas | 2020-04-05 | 2020-07-03 | 89 days *after* detected drift |
| Virginia | 2020-04-04 | 2020-05-29 | 55 days *after* detected drift |
| Washington | 2020-03-27 | 2020-06-26 | 91 days *after* detected drift |